# Fe$_3$O$_4$(110)-(1×3) Revisited: Periodic (111) Nano-Facets

Gareth S. Parkinson*, Peter Lackner, Oscar Gamba, Sebastian Maaß, Stefan Gerhold, Michele Riva, Roland Bliem, Ulrike Diebold and Michael Schmid

Institute of Applied Physics, TU Wien, Vienna, Austria

*parkinson@iap.tuwien.ac.at

**Abstract**

The structure of the Fe$_3$O$_4$(110)-(1×3) surface was studied with scanning tunneling microscopy (STM), low-energy electron diffraction (LEED), and reflection high energy electron diffraction (RHEED). The so-called one-dimensional reconstruction is characterised by bright rows that extend hundreds of nanometers in the [$\bar{1}$10] direction and have a periodicity of 2.52 nm in [001] in STM. It is concluded that this reconstruction is the result of a periodic faceting to expose {111}-type planes with a lower surface energy.

**Main Text**

Magnetite (Fe$_3$O$_4$) is a common material in the Earth's crust and plays an important role in geochemistry and corrosion [1; 2]. At room temperature Fe$_3$O$_4$ crystallizes in the inverse-spinel structure, and Fe cations occupy tetrahedrally (Fe$_{tet}$) and octahedrally (Fe$_{oct}$) coordinated interstices within a face-centred cubic lattice of O$^{2-}$ anions. Natural single crystals are typically octahedrally shaped and expose {111} facets, consistent with density functional theory (DFT)-based calculations that find (111) to be the most stable low-index surface [3; 4]. In recent years however, advances in synthesis have allowed the size and shape of Fe$_3$O$_4$ nanomaterial to be tailored with a view to enhancing performance in applications such as groundwater remediation, biomedicine, and heterogeneous catalysis [1; 5], and nanocubes and nanorods exposing {100} surfaces have been reported [6; 7]. To date, there have been no reports of Fe$_3$O$_4$ nanomaterial exhibiting {110} surfaces.

In this light it is perhaps unsurprising that the majority of studies aimed at uncovering the structure-function relationship of Fe$_3$O$_4$ surfaces have focussed on the (111) and (100) facets [2; 8; 9]. Nevertheless, there have been a handful of experimental studies of single crystals cut in the (110) direction [10-12], and Fe$_3$O$_4$(110) thin films have been successfully grown on MgO(110) [13-16]. In most situations a (1×3) reconstruction has been reported. In STM, the (1×3) surface has been shown to exhibit an unusual appearance with bright rows that extend for hundreds of nanometers in the [$\bar{1}$10] direction, and has been termed a 1-dimensional reconstruction. However, there is no reliable model for this surface structure. Fe$_3$O$_4$(110) has also been studied by theoretical methods, but these investigations did not consider the reconstruction [17-22], instead focussing on a comparison of bulk-like surface terminations with a (1×1) unit cell. In this paper we revisit the surface structure of Fe$_3$O$_4$(110)-(1×3) using STM, LEED and RHEED experiments, and conclude that the "1D reconstruction" reported previously is the result of periodic faceting to expose the {111} planes, presumably due to their lower surface energy.

The STM experiments were performed in an ultrahigh vacuum (UHV) system with connected vessels for preparation and analysis, with base pressures of 1×10$^{-10}$ mbar and 5×10$^{-11}$ mbar, respectively. A natural Fe$_3$O$_4$(110) single crystal (SurfaceNet GmbH) was prepared by cycles of 10 minutes sputtering (1 keV Ar$^+$ ions, ≈ 2 μA/cm$^2$) and subsequent annealing. The influence of the annealing temperature (varied between 400 and 900 °C) and environment (from UHV to an O$_2$ partial pressure of 10$^{-6}$ mbar) were systematically studied. A summary of the surfaces prepared is available in the supplementary information. The 1D reconstruction observed previously by Jansen et al [10] in STM images was omnipresent for all conditions, although the length of the 1D rows was maximised at 800 °C. XPS measurements were acquired with a non-monochromatized Al Kα source and a SPECS PHOIBOS 100 analyser with a pass energy of 90 eV. Temperatures were measured with a K-type thermocouple spot-welded near the sample plate and are underestimated by ≈ 50 °C. STM measurements were conducted using an Omicron μ-STM with electrochemically etched W tips in constant current mode. STM images were corrected for creep of the piezo scanner [23].

The best-quality $Fe_3O_4$(110)-(1×3) LEED pattern (Fig. 1b) was obtained after annealing at 800 °C, consistent with the previous work of Jansen et al [10]. The LEED spots are consistent with a periodicity of 0.3 nm and 2.5 nm in the [$\bar{1}$10] and [001] direction, respectively. A second set of LEED spots suggestive of an additional 0.6 nm periodicity along [$\bar{1}$10] direction are weakly visible, and we note that these spots appeared stronger for lower annealing temperatures. STM images of the $Fe_3O_4$(110) sample annealed at 800 °C reveal rows in the [$\bar{1}$10] direction (Fig. 1a), as observed previously by other groups [10; 12]. In overview images (Fig. 1a), the rows extend over hundreds of nanometers without break, and have a periodicity in the [001] direction of ≈ 2.52 nm (i.e. approximately three times the lattice parameter of 0.8396 nm). Steps on this surface have an apparent height of 0.3 nm, consistent with repeat distance of equivalent layers in the (110) direction (0.297 nm). At lower annealing temperatures (400-650 °C) the appearance of the surface is similar, but the rows feature more kinks and steps (see supplement, Fig. S1). At higher temperatures (900 °C) some missing sections appear in the rows (see supplement, Fig. S1).

Many attempts were made to image the (1×3) surface with atomic resolution while repeatedly preparing the sample, but no consistent structure was obtained. Indeed, it quickly became clear that such images were primarily dominated by the structure of the STM tip, because modification of the tip shape by $Ar^+$ sputtering and/or pre-scanning a clean Au(110) single crystal led to a different corrugation even on the same sample preparation. This experience suggests that the electronic corrugation along the rows is weak, while across the rows the surface is "sharper" than the STM tip. Consequently, for analysis we focus on the subset of images in which the corrugation across the rows is highest. One such image is shown in Fig 1c, together with a line profile in the [001] direction (Fig. 1d). The measured corrugation of the surface (which should be taken as a minimum value due to tip convolution effects [24]) is 0.46 nm, which is significantly larger than the distance between similar layers in the [110] direction (0.30 nm). Furthermore, the maximum slope angles measured are 30-35°, which is close to the angle of 35° expected between the (110) and (111) planes (blue line in Fig. 1d). Although not a direct measurement due to the convolution with tip shape [24], such a measurement does represent a minimum value for the real surface [25; 26]. Over the course of many experiments, a value in excess of 35° was never observed. A line profile along the top of the ridge (See Fig. S2) reveals an irregular structure, although it is common that two intensity maxima are 0.6 nm apart. This periodicity corresponds to the very weak LEED spots in Fig 1b.

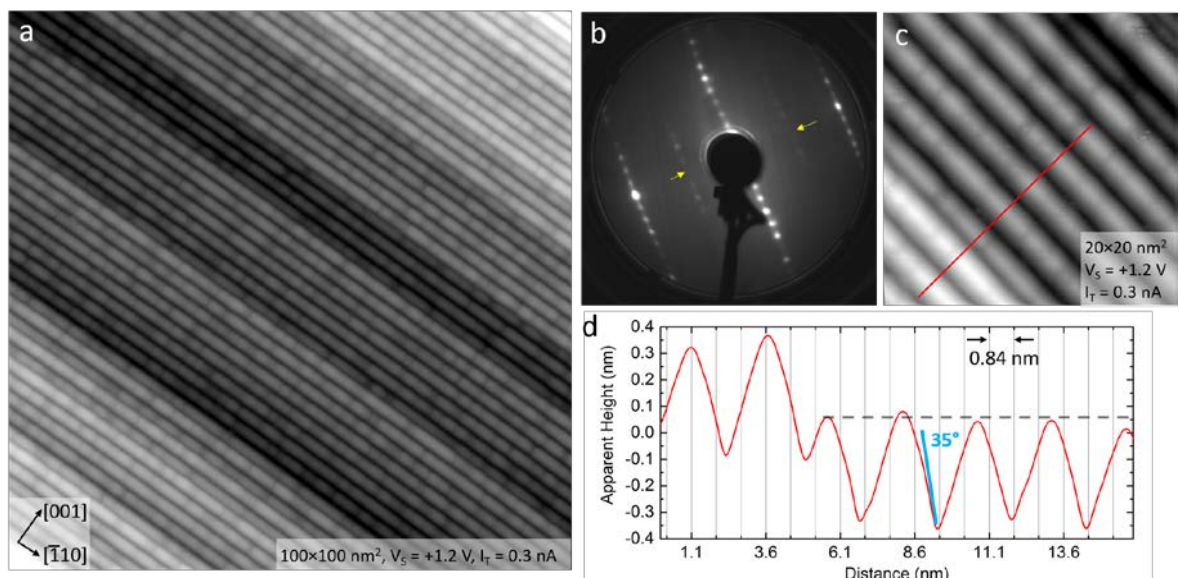

**Figure 1**: STM images of the $Fe_3O_4$(110) surface acquired after sputter/anneal cycles. (a) Overview image (100×100 nm$^2$, $V_S$ = +1.2 V, $I_T$ = 0.3 nA) showing bright rows that run in the [$\bar{1}$10] direction for hundreds of nm without break. Step edges have a height of 0.3 Å and are parallel to the rows. The periodicity in the [001] direction is 2.52 nm. (b) The (1×3) reconstruction observed in LEED. Weak spots due to an additional 0.6 nm periodicity are also visible, and exhibit extinctions due to a glide symmetry (yellow arrows). (c) Small-area STM image (20×20 nm$^2$, $V_S$ = +1.2 V, $I_T$ = 0.3 nA) including a step in the bottom left of the image. (d) Line profile acquired from the position of the red line in (c). The grid lines in the plot correspond to 0.84 nm, i.e. one bulk unit cell. Each peak is separated by three grid lines, and the registry shifts by ½ of one unit cell between alternate layers. The maximum slope measured by STM is ≈ 35°, in agreement with the proposed {111} facet model, and the measured corrugation of the ridge-trough structure (0.46 nm) is greater than the step height (0.3 nm).

To investigate whether the (1×3) reconstruction might be related to the formation of {111} nanofacets we performed RHEED experiments in a separate vacuum system (Fig. 2). The sample was prepared in the same way, and the presence of the (1×3) reconstruction was confirmed by in-situ LEED and STM (SPECS STM 150 Aarhus). Strong reflections were observed at an angle of 70° for all electron energies between 15 and 30 keV when the incident beam was aligned parallel to the [$\bar{1}$10] direction. This is the specular direction for {111} facets (70° = 2×35°). The high density of spots on the Laue circle of the RHEED images is related to the (1×3) superstructure. The large (2.52 nm) periodicity of this reconstruction results in many closely spaced spots that vary in intensity with electron energy.

On the basis of the experimental results shown in Figs. 1 and 2, we propose a model of the $Fe_3O_4$(110) surface based on {111} nanofacets. One possible realisation of such a structure is presented as Fig. 3 (green atoms are $Fe_{tet}$, blue atoms are $Fe_{oct}$, O is red). At the apex of the structure there is a row of $Fe_{oct}$ atoms, together with their neighbouring O atoms. In this respect our model is similar to that of Jansen et al. [10], who observed a 3 Å periodicity along the top of the rows with STM. Indeed, with an idealised STM tip one would expect to image the $Fe_{oct}$ atoms as a row of protrusions in empty-states images as they have density of states near $E_F$, and this is regularly achieved on the $Fe_3O_4$(100) surface. However, as mentioned above, we were unable to image this structure reproducibly despite repeated attempts, most likely due to the convolution of the surface structure with the tip morphology. Nevertheless, the structure is consistent with the (1×3) periodicity observed in LEED, and sometimes weakly observed in Fourier transforms of STM images. Note that O atoms are not typically imaged on $Fe_3O_4$ surfaces as they have little density of states in the vicinity of $E_F$ [8], and Fe$_{tet}$ have DOS at approximately 2 eV below $E_F$ [27]. STM images of the surface in filled states are also dominated by the ridge-trough structure and are indistinguishable from those in empty states.

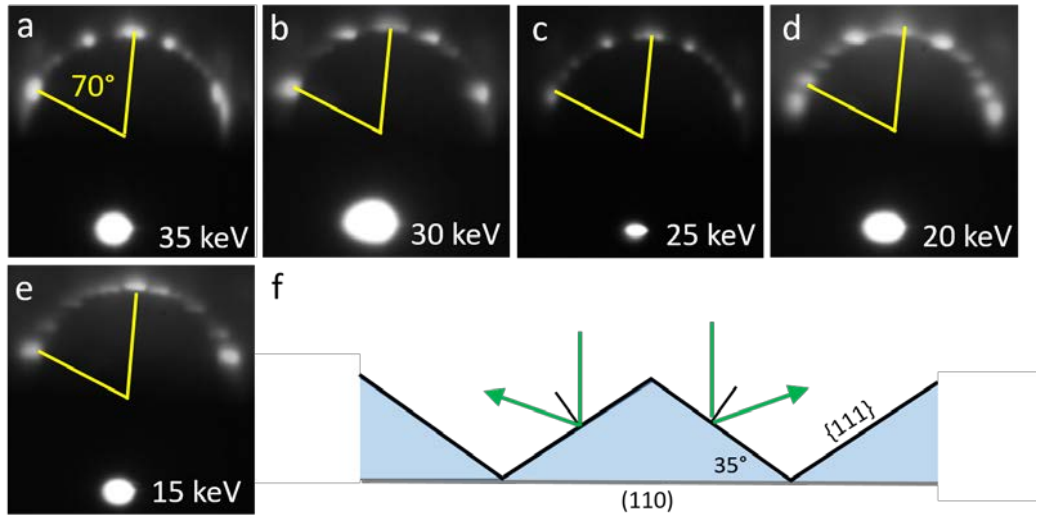

**Figure 2**: RHEED patterns acquired from the Fe$_3$O$_4$(110)-(1×3) surface for electron beam energies (a) 35 keV, (b) 30 keV, (c) 25 keV, (d) 20 keV and (e) 15 keV. For all energies there is significant scattered intensity at an angle of ≈70°. (f) The electron beam was aligned with the $\bar{1}$10 azimuth and scatters through an angle of 70° from the exposed {111} facets.

Where our model differs from that of Jansen et al. [10] is that they propose that the trough contains the subsequent Fe$_{oct}$-Fe$_{tet}$-O containing layer, and that the driving force for reconstruction is related to stoichiometry. We, on the other hand propose the driving force for reconstruction is anisotropy in the surface free energy. Such considerations drive faceting on open metal surfaces [28-30], where the exposure of close-packed surfaces is energetically favourable. On metal oxides, faceting has been observed for rocksalt compounds such as NiO(100) [31] and MnO [32], and attributed to a strong preference for the non-polar (100) surface [33]. Such logic cannot be applied to Fe$_3$O$_4$(110), because the unreconstructed Fe$_3$O$_4$(110) and Fe$_3$O$_4$(111) surfaces are both polar. Nevertheless, DFT+U calculations find a significant difference in the surface free energy after relaxation of the respective structures. For example, a recent paper [34] calculated a surface energy of 36 meV/Å$^2$ for the Fe$_{tet1}$ termination of Fe$_3$O$_4$(111), and proposed that large relaxations of the interlayer distances overcome the polarity issue. In contrast, a surface energy of ≈60 meV/Å$^2$ has been calculated for the fully relaxed Fe$_{oct}$-O termination of Fe$_3$O$_4$(110), and ≈80 meV/Å$^2$ for the Fe$_{oct}$-Fe$_{tet}$-O termination at the same O$_2$ chemical potential ($\mu_{O2}$ = -1.6 eV) [35]. Such a large energy difference explains how the ≈11% increase in surface area that results from the nano-faceting can be accommodated. It should be noted, however, that the surface energy depends strongly on the model structural model utilized. Interestingly, a termination of Fe$_3$O$_4$(110) with oxygen vacancies has been proposed to have a surface energy of ≈40 meV/Å$^2$ [35], which would in principle be competitive with the nanofaceted surface.

In Fig. 3, we have drawn {111} terminated facets in a roof-like structure consistent with the repeat distance of 3 unit cells. We have chosen to depict the facets with the so-called Fe$_{tet1}$ termination of Fe$_3$O$_4$(111), although a Fe$_{oct2}$ termination is also possible (both are present on Fe$_3$O$_4$(111) surfaces under reducing conditions [2; 36; 37]). An important feature of the model is that the ridge-trough height is 0.594 nm, which corresponds to the distance between two equivalent layers in the [110] direction. This is consistent with our STM measurements which show a ridge-trough distance greater than one repeat distance. For the structure drawn in Fig. 3 however, one would expect to see missing spots in LEED linked to glide symmetry of the outermost $Fe_{tet}$ atoms (green). We did observe such extinctions after annealing in $10^{-6}$ mbar $O_2$ at 650 °C, but the surface as prepared at 800 °C in UHV exhibits weak spots there (yellow arrows in Fig. 1b). Moreover, XPS measurements of the surface annealed with and without a partial pressure of $O_2$ (Fig. S3) reveal that the stoichiometry in the surface region changes with this treatment, although there is little difference visible in STM. This suggests that the atomic-scale structure of the ridge changes with the preparation conditions, although the preference for the {111} nanofacets does not.

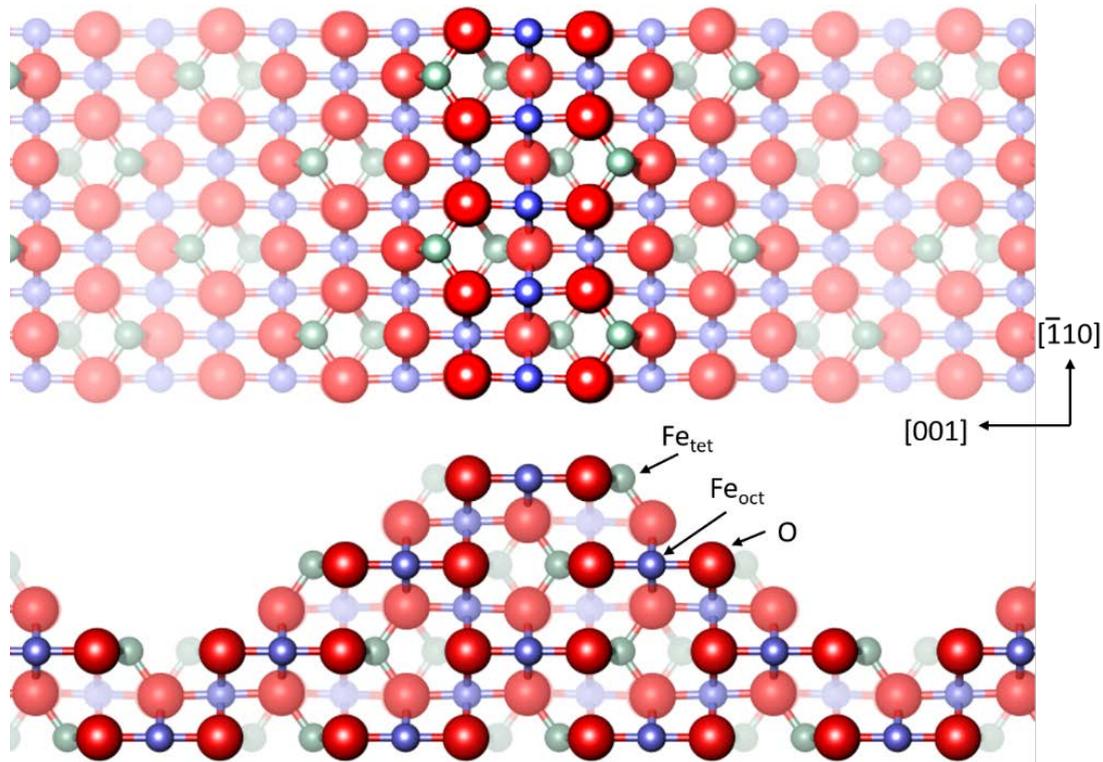

**Figure 3**: Top and side views of a proposed $Fe_3O_4$(111)-facet model of the $Fe_3O_4$(110)-(1×3) reconstruction. $Fe_{oct}$ atoms are blue, $Fe_{tet}$ atoms are green, and oxygen atoms are red. The {111} facets are drawn as exposing the $Fe_{tet1}$ termination of $Fe_3O_4$(111).

Finally, although the {111} nano-facets reconstruction of $Fe_3O_4$(110) dominates for single crystals prepared in UHV, it is possible that a $Fe_3O_4$(110) surfaces could be stabilized in a thin-film system if the metal-oxide interface contributes significantly to the surface energy. Epitaxial stabilization was recently proposed to explain the stability of the polar MnO(111) surface on Pd(100) [38]. If the surface could be stabilized it would likely be active for oxidation reactions, as observed for $Co_3O_4$(110) [39].

In summary, the experimental data presented here suggest that the $Fe_3O_4$(110) surface is unstable with respect to $Fe_3O_4$(111), consistent with theoretical predictions, and the lack of (110) facets on $Fe_3O_4$ nanomaterial. Nevertheless, the (1×3) surface is an extremely well-ordered 1D structure, and as such might provide a useful template for the growth of 1D nanostructures.

**Acknowledgements**

GSP, OG, and RB acknowledge funding from the Austrian Science Fund START prize Y 847-N20 and project number P24925-N20. RB and OG acknowledge a stipend from the Vienna University of Technology and Austrian Science Fund doctoral college SOLIDS4FUN, project number W1243. UD and SG acknowledge support by the European Research Council (ERC) Advanced Grant "OxideSurfaces". PL, MR, and MS were supported by the Austrian Science Fund (FWF) within SFB F45 "FOXSI".

**References**

[1]     R.M. Cornell, U. Schwertmann, The Iron Oxides, Wiley-VCH, 1996.
[2]     G.S. Parkinson, Surf. Sci. Rep. in press (2016).
[3]     D. Santos-Carballal, A. Roldan, R. Grau-Crespo, N.H. de Leeuw, Phys. Chem. Chem. Phys. 16 (2014) 21082-21097.
[4]     X. Yu, C.-F. Huo, Y.-W. Li, J. Wang, H. Jiao, Surf. Sci. 606 (2012) 872-879.
[5]     P. Tartaj, M.P. Morales, T. Gonzalez-Carreño, S. Veintemillas-Verdaguer, C.J. Serna, Adv. Mater 23 (2011) 5243-5249.
[6]     G. Gao, X. Liu, R. Shi, K. Zhou, Y. Shi, R. Ma, E. Takayama-Muromachi, G. Qiu, Cryst. Growth Des 10 (2010) 2888-2894.
[7]     M. Abbas, M. Takahashi, C. Kim, J. Nanopart. Res 15 (2012) 1-12.
[8]     R. Bliem, E. McDermott, P. Ferstl, M. Setvin, O. Gamba, J. Pavelec, M.A. Schneider, M. Schmid, U. Diebold, P. Blaha, L. Hammer, G.S. Parkinson, Science 346 (2014) 1215-1218.
[9]     W. Weiss, W. Ranke, Prog. Surf. Sci. 70 (2002) 1-151.
[10]    R. Jansen, V.A.M. Brabers, H. van Kempen, Surf. Sci. 328 (1995) 237-247.
[11]    O. Yuko, M. Seigi, T. Sakae, T. Eiko, H. Kazunobu, Jpn. J. Appl. Phys 37 (1998) 4518.
[12]    G. Maris, O. Shklyarevskii, L. Jdira, J.G.H. Hermsen, S. Speller, Surf. Sci. 600 (2006) 5084-5091.
[13]    R. Jansen, H. van Kempen, R.M. Wolf, J. Vac. Sci. Tec. 14 (1996) 1173-1175.
[14]    R.G.S. Sofin, S.K. Arora, I.V. Shvets, Phys. Rev. B 83 (2011) 134436.
[15]    M. Gabriela, J. Lucian, G.H.H. Jan, M. Shane, M. Giuseppe, V.S. Igor, S. Sylvia, Jpn. J. Appl. Phys 45 (2006) 2225.
[16]    G. Maris, L. Jdira, J.G.H. Hermsen, S. Murphy, G. Manai, I.V. Shvets, S. Speller, Magnetics, IEEE Transactions on 42 (2006) 2927-2929.
[17]    Y.-l. Li, K.-l. Yao, Z.-l. Liu, Front. Phys. China 2 (2007) 76-80.
[18]    T. Yang, X.-d. Wen, J. Ren, Y.-w. Li, J.-g. Wang, C.-f. Huo, Journal of Fuel Chemistry and Technology 38 (2010) 121-128.
[19]    X. Yu, Y. Li, Y.-W. Li, J. Wang, H. Jiao, J. Phys. Chem. C 117 (2013) 7648-7655.
[20]    U. Aschauer, A. Selloni, J. Chem. Phys. 143 (2015) 044705.
[21]    X. Yu, X. Zhang, S. Wang, Appl. Surf. Sci. 353 (2015) 973-978.
[22]    Y.L. Li, K.L. Yao, Z.L. Liu, Surf. Sci. 601 (2007) 876-882.
[23]    J.I.J. Choi, W. Mayr-Schmölzer, F. Mittendorfer, J. Redinger, U. Diebold, M. Schmid, J. Phys.: Condens. Matter 26 (2014) 225003.
[24]    M. Stedman, J. Microsc 152 (1988) 611-618.
[25]    J. de la Figuera, Z. Novotny, M. Setvin, T. Liu, Z. Mao, G. Chen, A.T. N'Diaye, M. Schmid, U. Diebold, A.K. Schmid, G.S. Parkinson, Phys. Rev. B 88 (2013) 161410.
[26]    H. Wang, W. Chen, R.A. Bartynski, P. Kaghazchi, T. Jacob, J. Chem. Phys. 140 (2014) 024707.
[27]    A. Yanase, N. Hamada, J. Phys. Soc. Jpn 68 (1999) 1607.
[28]    T.E. Madey, W. Chen, H. Wang, P. Kaghazchi, T. Jacob, Chem. Soc. Rev. 37 (2008) 2310-2327.
[29]    Q. Chen, N.V. Richardson, Prog. Surf. Sci. 73 (2003) 59-77.
[30]    G. Binnig, H. Rohrer, C. Gerber, E. Weibel, Surf. Sci. 131 (1983) L379-L384.
[31]    C. Mocuta, A. Barbier, G. Renaud, Y. Samson, M. Noblet, J. Magn. Magn. Mater. 211 (2000) 283-290.
[32]    K. Meinel, M. Huth, H. Beyer, H. Neddermeyer, W. Widdra, Surf. Sci. 619 (2014) 83-89.

[33]     G. Jacek, F. Fabio, N. Claudine, Rep. Prog. Phys. 71 (2008) 016501.
[34]     J. Noh, O.I. Osman, S.G. Aziz, P. Winget, J.-L. Brédas, Chem. Mater 27 (2015) 5856-5867.
[35]     Y.L. Li, K.L. Yao, Z.L. Liu, Surf. Sci. 601 (2007) 876-882.
[36]     P. Dementyev, K.-H. Dostert, F. Ivars-Barceló, C.P. O'Brien, F. Mirabella, S. Schauermann, X. Li, J. Paier, J. Sauer, H.-J. Freund, Angewandte Chemie International Edition 54 (2015) 13942–13946.
[37]     C. Lemire, R. Meyer, V.E. Henrich, S. Shaikhutdinov, H.J. Freund, Surf. Sci. 572 (2004) 103-114.
[38]     F. Allegretti, C. Franchini, V. Bayer, M. Leitner, G. Parteder, B. Xu, A. Fleming, M.G. Ramsey, R. Podloucky, S. Surnev, F.P. Netzer, Phys. Rev. B 75 (2007) 224120.
[39]     X. Xie, Y. Li, Z.-Q. Liu, M. Haruta, W. Shen, Nature 458 (2009) 746-749.